\documentclass[twocolumn,showpacs,preprintnumbers,amsmath,amssymb]{revtex4}

\usepackage{graphicx}
\usepackage{dcolumn}
\usepackage{bm}

\begin{document}

\title{Voltage-Controlled Spin Selection in a Magnetic Resonant Tunnelling Diode.}

\author{A. Slobodskyy}
\author{C. Gould}%
\author{T. Slobodskyy}%
\author{C.R. Becker}%
\author{G. Schmidt}%
\author{L.W. Molenkamp}%

\affiliation{%
Physikalisches Institut (EP3), Universit\"{a}t W\"{u}rzburg, Am
Hubland, D-97074 W\"{u}rzburg, Germany
}%

\date{ March 7, 2003.}

\begin{abstract}
We have fabricated all II-VI semiconductor resonant tunneling
diodes based on the (Zn,Mn,Be)Se material system, containing
dilute magnetic material in the quantum well, and studied their
current-voltage characteristics. When subjected to an external
magnetic field the resulting spin splitting of the levels in the
quantum well leads to a splitting of the transmission resonance
into two separate peaks. This is interpreted as evidence of
tunneling transport through spin polarized levels, and could be
the first step towards a voltage controlled spin filter.
\end{abstract}

\maketitle

The physics governing spin injection and detection in
semiconductor structures is now well understood
\cite{Schm,Rash,Fert}. The problems associated with the impedance
mismatch between a magnetic layer and the semiconductor can be
overcome, e.g. by using dilute magnetic semiconductor injectors
\cite{Fied,Ohno,Schm1}, or by fitting metallic magnetic contacts
with tunnel barriers \cite{Motsnyi}. However, these options can
only be utilised to transfer majority spin from the magnetic
material into the non-magnetic layer and, similar to the situation
in magnetic metallic multilayers, contacts with different shape
anisotropy must be used when the direction of the spin of the
electrons in the semiconductor is to be detected. Rather than
having to use an external magnetic field to switch the contact
magnetisation, it would be very desirable to have devices where
the spin character of the injected or detected electrons could be
voltage selected. Here we report on the successful operation of a
magnetic resonant tunnelling diode (RTD), which we hope will prove
useful for voltage controlled spin polarised injection and
detection \cite{DiVi}.

\begin{figure}
\centerline{\includegraphics{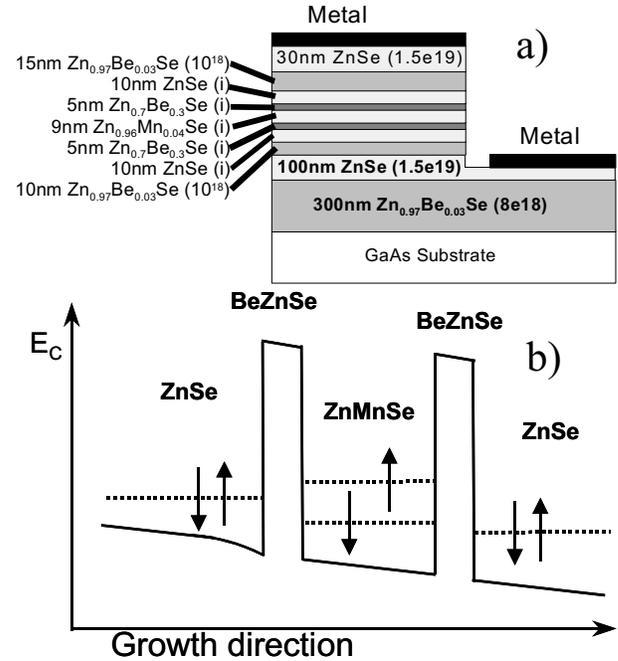}} \caption{\label{mbe} a)
Layer structure of the device and b) schematic view of resonance
tunnel diode band structure under bias.}
\end{figure}

The idea behind the RTD scheme is fairly straightforward but its
realisation was previously hampered by material issues
\cite{Gruber}. Since the well is made of magnetic material, the
energy levels in the well will split into spin-up and spin-down
states, as sketched in Fig.~\ref{mbe}(b). By selectively bringing
the spin-up or spin-down state into resonance, one can
dramatically increase the transmission probability of the desired
spin species.

In this paper, we describe our experimental investigations of an
all II-VI semiconductor RTD based on ZnBeSe, and with a ZnMnSe
dilute magnetic semiconductor (DMS) quantum well. In the presence
of a (constant) magnetic field, the DMS exhibits a giant Zeeman
splitting that, at low temperatures, leads to an energy splitting
of the Zeeman levels in the conduction band of about 15 meV at
fields of one to two Tesla. Our samples show typical RTD-like
current-voltage (I-V) characteristics with peak to valley ratios
of over 2.5 to one. As a function of applied magnetic field, the
transmission resonance of the I-V curve splits into two peaks with
a splitting corresponding to the separation of the energy levels
in the well.

The investigated II-VI semiconductor heterostructures were grown
by molecular-beam epitaxy on insulating GaAs substrates. The
active region of the device consists of a 9 nm thick undoped
Zn$_{.96}$Mn$_{.04}$Se quantum well, sandwiched between two 5 nm
thick undoped Zn$_{.7}$Be$_{.3}$Se barriers. The complete layer
structure is shown in Fig.~\ref{mbe} along with a schematic of the
potential energy profile of the double-barrier structure under
bias. The bottom 300 nm of Zn$_{.97}$Be$_{.03}$Se, highly n-type
doped with iodine to a concentration of n=8*10$^{18}$cm$^{ - 3}$,
as well as the 15 and 10 nm thick Zn$_{.97}$Be$_{.03}$Se layers
with n=1*10$^{18}$ cm$^{ - 3}$ in the injector and collector
contain 3{\%} Be in order to be lattice matched to the GaAs
substrate. The barrier layers are clearly not lattice matched to
the substrate, but are sufficiently thin to be grown as fully
strained epitaxial layers. The heterostructure was patterned into
square mesas with side of 100, 120 and 150 $\mu $m by optical
lithography with positive photoresist followed by metal
evaporation and lift off.

Special care must be taken in order to obtain good ohmic contacts
to the II-VI semiconductor. For the top contact, this can be
achieved by in-situ growth of a metal contact layer consisting of
10 nm Al to achieve good contact, 10 nm Ti as a diffusion barrier,
and 30 nm Au to avoid oxidation. This procedure reliably yields
contact resistivities of the order of 10$^{ - 3} \Omega
$.cm$^{2}$. Since the bottom contact can only be fabricated after
processing, it must rely on an ex-situ technique where the contact
resistivity is typically 1-3 orders of magnitude higher. Its
resistance is kept reasonably low by a combination of the
incorporation of a highly n doped 100 nm ZnSe contact layer in the
heterostructure, and the use of a relatively large (500$^{2} \mu
$m$^{2})$ Ti-Au contact pad.

The samples were inserted into a $^{4}$He bath cryostat equipped
with a 6 Tesla superconducting magnet, and were investigated using
standard low noise electrical characterization techniques.
Precautions were taken to prevent problems associated with the
measurement circuit going into oscillations. A stabilised voltage
source was used to apply bias to the circuit, which consists of
the RTD, a 33 Ohm reference resistance in series and a 40 Ohm
resistor in parallel. Such a setup is known to prevent
bi-stability of the circuit in the region of negative differential
resistance \cite{Leadb}. By measuring the voltage drop over the
RTD and the reference resistor as the bias voltage is swept, I-V
curves of the RTD can be extracted. The absence of charging in the
device was confirmed by comparing I-V curves with different sweep
direction \cite{Martin}.

We studied devices with Mn concentrations of 4{\%} and 8{\%} in
the quantum well layer. The layers thicknesses of the 4{\%} sample
are given in Fig.~\ref{mbe}, whereas those for the 8{\%} Mn sample
are all 6{\%} thinner. For structures with 4{\%} Mn in the well,
the first resonance for positive bias occurs at 105 mV and has a
peak to valley ratio of 2.5 whereas for the 8{\%} Mn sample the
resonance is at 127 mV with a peak to valley ratio 2.25. The size
of the mesas had no effect on the position and strength of the
resonance.

\begin{figure}
\centerline{\includegraphics{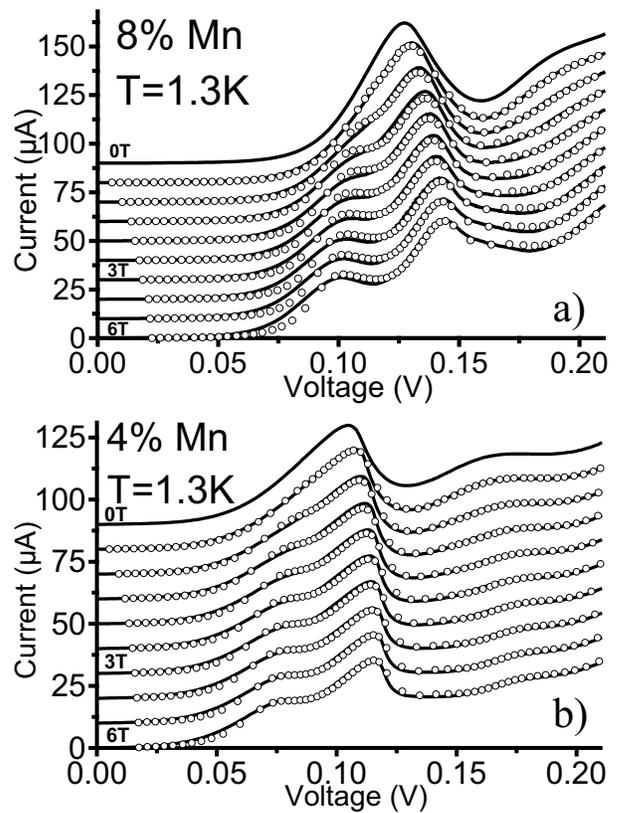}} \caption{\label{feild}
Experimental (lines) and modeled (circles) I-V curves for a
resonance tunnel diode with a) Zn$_{.92}$Mn$_{.08}$Se and b)
Zn$_{.96}$Mn$_{.04}$Se in the quantum well. Curves taken in 0.5T
intervals from 0 to 3T and in 1T intervals from 3T to 6T.}
\end{figure}

I-V characteristics of the RTDs were measured in different
magnetic fields in the range from 0 to 6T applied either
perpendicularly to, or in the plane of the quantum well. The
results for perpendicular magnetic fields are presented in
Fig.~\ref{feild} (lines). For clarity, subsequent curves are
offset by 10 $\mu $A. It is clear from the figure that the
resonance is split into two parts and that the splitting grows as
a function of magnetic field. At 6 Tesla, the separation between
the maxima of the split peaks is 36.5 mV and 42 mV for 4{\%} and
8{\%} Mn samples respectively. The broader feature which can be
seen most prominently in the zero field curve at approximately 165
mV (180 mV) for the 4{\%} (8{\%}) of Mn samples is an LO phonon
replica \cite{Leadb}.

To explain the magnetic field-induced behaviour of the resonance,
we develop a model based on the Giant Zeeman splitting for the
spin-split levels in the DMS quantum well. First, we extract the
series contact resistance of the RTD from the measured I-V curve
at zero magnetic field \cite{Martin,Lerch}. Then we assume that
each of the two spin split levels have the same conductance, and
that each therefore carries half of the current in the device.

We neglect the slight relative change of barrier height caused by
the change in energy of the levels, and assume that the
conductivity of each level is independent of magnetic field in the
sense that for the same alignment between the emitter Fermi level
and the well level, the conductivity will be the same. In other
words, the conductivity of each level as a function of applied
voltage in the presence of a magnetic field can be given by a
simple translation of the zero field curve by a voltage
corresponding to the energy shift of the pertinent spin level in
the well.

After translation, we add the conductivity contributions of both
the spin-up and the spin-down curves, and reinsert the series
resistance to yield a modelled I-V curve at a given magnetic
field. By comparing this I-V curve with the actual experimental
curve at 6 Tesla, we determine an optimal value of 330 Ohm (900
Ohm) for the series resistance for the 8{\%} (4{\%}) Mn sample.
Using this value of the series resistance for all modelled curves,
and fitting the modelled curves at each field to the experimental
ones, we extract the voltage splitting of the levels $\Delta $V as
a function of magnetic field. Fig.~\ref{feild} shows the modelled
I-V curves as the circles, which compare very well with the
experimental data presented as the solid lines in the figure.

\begin{figure}
\centerline{\includegraphics{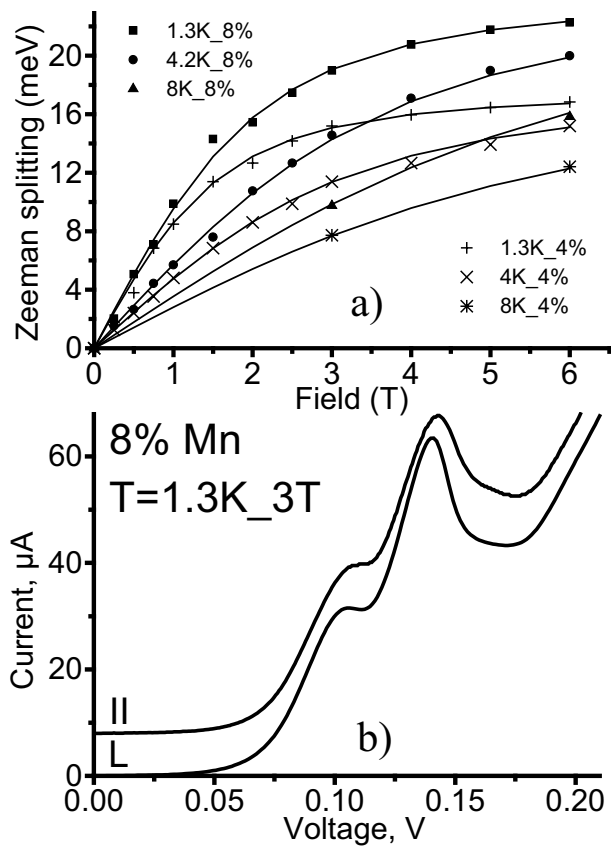}} \caption{\label{split} a)
Giant Zeeman splitting versus magnetic field for both samples. The
lines come from a Brillouin like description of the well levels.
The symbols represent the splitting in the peak positions
extracted from the experimental data. b) I-V curves for sample
with 8{\%} Mn, under 3T in plane (\textsf{II}) or perpendicular
(\textsf{L}) magnetic field.}
\end{figure}

We now compare the values of $\Delta $V extracted from this
fitting procedure to the expected behaviour of the spin levels in
the wells. The spin level splitting \textit{$\Delta $E} of the DMS
as a function of magnetic field \textit{B} is given by a modified
Brillouin function \cite{Gaj}:

\[
\Delta E = N_0 \alpha \,x\,s_0 B_s \left( {s\,g\;\mu _B B / k_B (T
+ T_{eff} )} \right),
\]

\noindent where $N_{0}$\textit{$\alpha$} is the s-d exchange
integral, $x, s, $and $g$ are the manganese concentration,
manganese spin, and g-factor respectively, and \textit{$\mu
$}$_{B}$ is the Bohr magnetron. $B_{s}$ is the Brillouin function
of spin \textit{s}. $s_{0}$ and $T_{eff}$ are, respectively, the
effective manganese spin and the effective temperature. These
phenomenological parameters are needed to account for Mn
interactions. Taking established values of $T_{eff }$= 2.24 K,
$s_{0 }$= 1.13 \cite{Yu} for $x$=8{\%} ($T_{eff }$= 1.44 K, $s_{0
}$= 1.64 for $x$=4{\%}) and $N_{0}$\textit{$\alpha $} 0.26 eV
\cite{Tward} from the literature, we obtain values for
\textit{$\Delta $E} which are plotted as the solid lines in
Fig.~\ref{split}(a) for temperatures of 1.3, 4.2, and 8K and for
Mn concentrations of 4{\%} and 8{\%}. These curves are compared
with the values of \textit{$\Delta $E} extracted from experiment
(symbols in the same figure) for measurements taken at the
respective temperatures. It is important to note that in order to
correctly fit the amplitude to the Brillouin function, the
measured values of $\Delta $V must be divided by a lever arm of
2.1. The agreement between the magnetic field dependence of the
experimental values and the Brillouin function is remarkable,
suggesting that our model of two spin level splitting in a
magnetic field captures the essential character of the device.
More over an identical value of the lever arm is found for both
the 4{\%} and 8{\%} Mn samples.

The existence of a lever factor between the experimental voltage
splitting and the theoretical energy splitting in the well is a
well-known feature of RTDs. It occurs because only part of the
voltage applied to the device is dropped over the first barrier,
and thus effective in determining the alignment condition for the
resonance. Furthermore, extracting the lever factor for our diode
by comparing the observed voltage splitting in the \textit{B}=0
I-V curve between the first resonance and the phonon replica to
the known LO phonon energy of ZnSe (31.7 meV) \cite{Landolt} also
yields a lever arm of around 2.

In Fig.~\ref{split}(b) we compare measurements at 3T with magnetic
field in the plane of, and perpendicular to the quantum well. The
curves are offset for clarity. Evidently, the peaks in the I-V
curve for in-plane magnetic field are broader than those for
perpendicular field and are shifted slightly towards higher bias
voltages. This effect is known from GaAs based RTDs \cite{Amor}
and can be explained in terms of momentum conservation. For
magnetic field in plane (perpendicular to the motion of the
electrons), the Lorentz force will give the electrons an in-plane
momentum. The conservation of in-plane momentum during the
tunnelling process forces electrons to tunnel into finite momentum
states of the quantum well, which are at higher energy than the
zero momentum ground state. Furthermore, the spread in in-plane k
vectors leads to a broadening of the tunnelling resonance
\cite{Amor,Davies}. From the figure, it is obvious that the
splitting of the resonance peak is similar in both orientations of
magnetic field. This is due to the isotropic g-factor in the DMS.
This observation directly rules out any explanation of the peak
structure as resulting from Landau level splitting in the quantum
well.

\begin{figure}
\centerline{\includegraphics{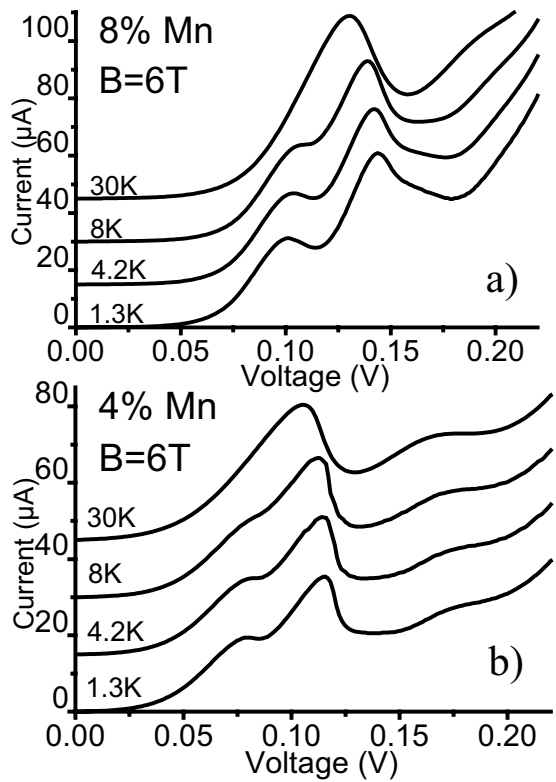}}
\caption{\label{temperature} Temperature dependence of the I-V
curves of the first resonance, shown for each of the diodes.}
\end{figure}

Temperature dependent measurements of the 6T I-V curves are
presented in Fig.~\ref{temperature}, again with the curves offset
for clarity. Qualitatively, an increase in temperature has a
similar effect on the I-V curves as a reduction of the magnetic
field (cf. Equ. 1). The peaks move closer together and eventually
merge back into a single peak. On the other hand, the zero field
I-V curves are practically temperature independent in the range
from 1.3 Kelvin to 30 Kelvin. The reduction of the splitting as a
function of field can be anticipated from the results of
Fig.~\ref{split}, and is simply a manifestation of the temperature
dependence of the giant Zeeman splitting of the DMS. This suggests
that any operational limit imposed on the device by temperature is
purely a function of the material in the well, and that no
inherent limit from the tunnelling process is detected.

In summary, we have presented results of all II-VI semiconductor
RTDs based on the (Zn,Be)Se system, with magnetic impurities (Mn)
in the quantum well. A strong splitting of quantum well resonance
is observed as a function of magnetic field, which originates from
the Giant Zeeman splitting of the spin levels in the dilute
magnetic semiconductor quantum well. An intuitive model that
simulates the magnetic field dependence of the I-V characteristic
of the device is discussed, and shows good agreement with the
experiment. The results therefore demonstrate the possibility of
devices based on tunnelling through spin resolved energy levels.
Experiments aiming to measure the spin polarisation of the current
flowing through such a device are currently underway.

\begin{acknowledgments}
The authors would like to thank G. Austing, H. Buhmann, L. Eaves,
A. Gr\"{o}ger and J. Wang for useful discussions and V. Hock for
sample fabrication, as well as DFG (SFB 410), BMBF, the DARPA
Spins program and ONR for financial support.
\end{acknowledgments}

\end{document}